# INCREASE OF ORGANIZATION IN COMPLEX SYSTEMS

Georgi Yordanov Georgiev*, Michael Daly, Erin Gombos, Amrit Vinod, Gajinder Hoonjan

*Abstract* — Measures of complexity and entropy have not converged to a single quantitative description of levels of organization of complex systems. The need for such a measure is increasingly necessary in all disciplines studying complex systems. To address this problem, starting from the most fundamental principle in Physics, here a new measure for quantity of organization and rate of self-organization in complex systems based on the principle of least (stationary) action is applied to a model system - the central processing unit (CPU) of computers. The quantity of organization for several generations of CPUs shows a double exponential rate of change of organization with time. The exact functional dependence has a fine, S-shaped structure, revealing some of the mechanisms of self-organization. The principle of least action helps to explain the mechanism of increase of organization through quantity accumulation and constraint and curvature minimization with an attractor, the least average sum of actions of all elements and for all motions. This approach can help describe, quantify, measure, manage, design and predict future behavior of complex systems to achieve the highest rates of self organization to improve their quality. It can be applied to other complex systems from Physics, Chemistry, Biology, Ecology, Economics, Cities, network theory and others where complex systems are present.

All authors are at the Department of Natural Sciences – Physics and Astronomy; Assumption College, Worcester MA, 01609, USA *Corresponding author: G.Y. Georgiev (phone: 508-767-7251; fax: 508-767-7241; ggeorgie@assumption.edu; georgi@alumni.tufts.edu)

## I. INTRODUCTION

IN the field of complex systems the process of progressive development is understood as a continuous improvement through self-organization. New structures, rules and laws in systems emerge determining new levels of organization. But, how is organization defined and how it and the rate of self-organization are to be measured and quantified? What is its mechanism and what is the potential for further self-improvement in complex systems. What are the limits of that improvement? Our society is a complex system, therefore the answers to those questions are vital and will help us better manage our economy, education and other systems, all infrastructure and communications networks. It will also help us understand better the physical, chemical and biological systems.

To answer the above questions, we apply a new measure for quantity of organization and rate of self-organization [1,2] based on the principle of least (stationary) action. Systems can be represented as networks. In [1] quantity of organization α is defined as the reciprocal of average number of quanta of action per one edge for the motions of the elements.

The principle of least action is the one "from which all other principles naturally flow" [3]. All branches of physics and all conservation laws have been derived from it. Recently it has been applied widely to networks and complex systems, like in network theory [4-6] and path integral approaches to stochastic processes and networks [7]. Samples of some other applications are by Annila and Salthe for natural selection [8] and Devezas for technological change [9]. Some of the other important measures and methods used in complex systems research are presented by Chaisson [10], Bar-Yam [11], Smart [12], Vidal [13] and Gershenson and Heylighen [14]. This list is not exhaustive. Some of these established measures use information, entropy or energy to describe complexity, while a fundamental quantity of physical action is used in this work to describe degree of organization through efficiency.

## II. METHOD

The principle of least action has been defined for a complex developing system which is represented as a network with n elements crossing m edges in [1] as:

$$\delta \sum_{i=0}^{n}\sum_{j=0}^{m} I_{ij} = \delta \sum_{i=1}^{n}\sum_{j=1}^{m} \int_{t_1}^{t_2} L_{ij} dt = 0. \quad (1)$$

Where $\delta$ is infinitesimally small variation in the action integral $I_{ij}$ of the j[th] crossings between the nodes (unit motion) of the i[th] element and $L_{ij}$ is the Lagrangian for that motion. $n$ represents the number of elements in a system, m the number of motions and $t_1$ and $t_2$ are the initial and final times of each motion. The double sum is the sum of all actions of all elements n for their motions m between nodes of a complex network. For example, a unit motion for electrons on a computer chip is the one necessary for one computation. For a computer network, such as internet, it is the transmission of one bit of information. In a chemical system it is the one for one chemical reaction. The state of zero variation of the total action for all motions is the one to which any system is naturally driven. Open systems never achieve this least action state because of the constant changes that occur in them, but are always tending toward it. In some respect one can consider this attractor state to be one of dynamical action equilibrium. Using the quantity of action one can measure how far the system is from this equilibrium and can distinguish between the organizations of two systems, both of which are equally close to equilibrium.

Organization, α, is defined as inversely proportional to the average action per one element and one motion [1,2]:

$$\alpha = \frac{hnm}{\sum_{i=0}^{n}\sum_{j=0}^{m} I_{ij}}. \qquad (2)$$

Where h is the Planck's constant. The meaning of organization is that it is inversely proportional to the number of quanta of action per one motion of one element in a system. This definition is for a system in which its elements can be approximated to be identical. It is the efficiency of physical action. The time derivative of α is the rate of progressive development of a complex system.

The domain of the function is between zero, for totally disorganized system, i.e. one at maximum entropy, to infinity, for a system where many edges can be crossed using one quantum of action or in quantum processes where particles take simultaneously an infinite number of paths from one node to another.

This definition can be applied to the organization of any complex system. Systems self-organize to decrease the average action per element per unit motion. This lowest action state is the attractor for the continuous self-organization and evolution of a dynamical complex system. Constraints increase this average action and constraint minimization by the elements is a basic mechanism for action minimization. Increase of quantity of elements in a network, leads to faster constraint minimization through grouping, decrease of average action per element and motion and therefore accelerated rate of self-organization. Progressive development, as self-organization, is a process of minimization of action.

III. MECHANISM OF SELF-ORGANIZATION

*A. One element and one constraint*

Consider the simplest possible part of a network: one edge, two nodes and one element moving from node 1 to node 2. Let's consider case (I) when there is no constraint for the motion of the element. It crosses the path between nodes 1 and 2 along the shortest line – a geodesic. Now consider case (II) when there is one constraint placed between nodes 1 and 2 and the shortest path of the element in this case is not a geodesic. If the path is twice as long in the second case, if the kinetic energy of the element is the same as in case (I) and no potentials are present, then the time taken to cross between nodes 1 and 2 is twice as long. Therefore the action in case (II) is twice than the action in case (I). When we substitute these numbers in the expression for organization α (eq. 2), where n=1, one element, and m=1, one crossing between two nodes, then the denominator which is just the action of the element for that motion will be twice as large in the second case and therefore the result for the amount of organization is a half as compared to the first case.

*B. Many elements and constraints*

Now consider an arbitrary networks consisting of three, ten, thousands, millions and billions of nodes and edges, populated by as many elements and constraints, where the paths of the elements cross each other. The optimum of all of the constraints', nodes', edges' and elements' positions and the motions of the elements is the minimum possible action state of the entire system, providing a numerical measure for its organization. Notice that action is not at an absolute possible minimum in this case, but at a higher, optimal value. Action would be at its absolute minimum only in a system without any constraints on the motion of its elements, which is not the case in complex systems and networks. Nevertheless, action is at a minimum compared to what it will be for all other arrangements of nodes, elements and constraints in the system that are less organized. When we consider an open dynamical system, where the number and positions of nodes, edges, elements and constraints constantly changes, then this minimum action state is constantly recalculated by the system. It is an attractor state which drives the system to higher level of organization and this process can continue indefinitely, as long as the system exists. Achieving maximum organization is a dynamical process in open complex systems of constantly recalculating positions of nodes, edges, elements and constraints for a least action state and preserving those positions in a physical memory of the organization of the system.

When elements interact with constraints they apply force to minimize them, lowering their action for the next cycle. With the increase of quantity in a system, several elements can group on the same constraint to minimize it for less time. Decreased average action makes a system more stable, by lowering the energy needed for each motion. High average action, in disorganized system destabilizes it and above some limit it falls apart. Therefore a system with low enough average action can increase its quantity within limits of stability. Quantity and level of organization are proportional. If the quantity becomes constant, then the organization will reach a least action state and stop increasing. For continued self-organization an increase of the quantity is necessary. Quantity and level of organization of a system are in an accelerating positive feedback loop, ensuring unlimited increase of the level of organization in a system, unless it is destroyed by external influence, like limited resources, huge influx of energy, force impact, change in the conditions, etc.

Consider a networks populated by elements and constraints. The optimum of all of the constraints', nodes', edges' and elements' positions and the motions of the elements is the minimum possible action state of the entire system, providing a numerical measure for its organization. The average action per element and edge is at a minimum compared to what it will be for all other arrangements of nodes, elements and constraints in the system that are less organized. When we consider an open dynamical system, where the number and positions of nodes, edges, elements and constraints constantly changes, then this minimum action state is constantly recalculated by the system. It is an attractor state which drives the system to higher level of organization and this process can continue indefinitely, as long as the system exists. Achieving maximum organization is a dynamical process in open complex systems of constantly recalculating positions of nodes, edges, elements and constraints for a least action state and preserving those positions in a physical memory of the organization of the system.

IV. DATA ANALYSIS AND RESULTS

In this work, the first model system that we applied the above definition to is the core processing unit (CPU) of computers, single core processors only, which are steadily becoming more complex and better organized. A unit of motion of CPUs elements is defined to be the one necessary for one computation. When the CPUs are modeled as networks, the nodes are the calculations, the edges are the distance travelled by the electrons to complete them, and the constraints are coming from the curvature of the path and friction. Using published data we calculated the quantity of organization for several generations of CPUs using the formula for α in [1].

To calculate α, the potential energy of the electrons was taken to be constant. The lagrangian was then calculated only through the kinetic energy. The data for Million Instructions per second (MIPS) for each processor was divided by the thermal design power and multiplied by the table value of the Planck's constant, to solve for α. This takes into account the energy lost in dissipation.

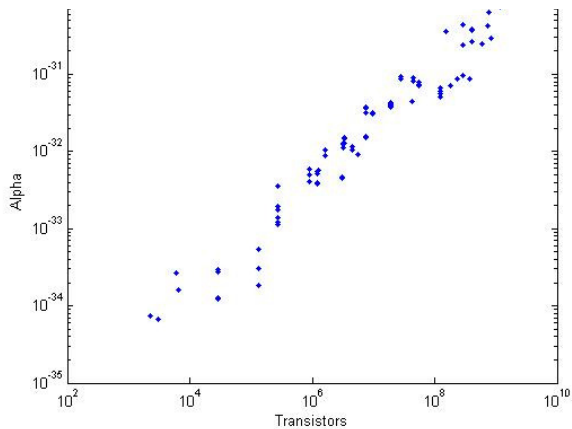

Fig. 1. α vs. transistor count since. On this log-log plot the organization increases proportional to the amount of transistors.

As shown on Fig. 1, the organization (quality) increases as a function of the transistor count (quantity), which in our view is a general property of growth of complex systems. The increase of quantity allows and causes the increase of quality, and the increase of quality keeps the system intact and allows and causes the increase of the quantity as well. This positive feedback functional dependence leads to accelerating, super-exponential growth of complexity.

We measured a double exponential rate of change of organization of CPUs with time and its exact functional dependence. The rate of increase is a double exponential and not a smooth curve, but increases with logistic steps which agrees with the expected "punctuated equilibrium", technology S-curves and related theories of increase of complexity and self-organization. Since 1971 we observe four of those steps. We fit each of those steps to a logistic equation and look for trends in the parameters of each step. Also, we find the frequencies of the steps by a Fourier analysis of the data. Using the principle of least action allows us to explain the mechanism of constraint minimization behind the numerical increase of organization with an attractor, the least sum of actions of all elements in a complex system.

V. CONCLUSIONS

The principle of least action for a networked complex system (eq. 1) drives self-organization in complex systems and the average action is the measure of degree to which they approach this least action state. Actions that are less than their alternatives are self-selected. Progressive development, as self-organization, is a process of minimization of action. In open systems there is a constant change of the number of elements, constraints and energy of the system and the least action state is different in each moment. The process of self-organization of energy, particles, atoms, molecules, organisms, to the today's society is a process of achieving a lower action state, with the least action as a final state. The laws of achieving this least action state are the laws of self-organization. The least possible action state is the limit for organization when time is infinite and all elements in the universe are included.

The state of nodes, edges, constraints and elements that determines the action for one motion in a system is its organization. With its measure α (eq. 2) we can compare any two systems of any size and the same system at two stages of its development. It distinguishes between systems with two different levels of organization and rates of self-organization and is normalized for their size.

Our conclusions from the data analysis for CPUs are that the numerical measure for α can serve as a good descriptor of organization and that it provides insights in the mechanisms and processes leading the higher levels of organization. The overall rate of organization and its fine structure are revealed. The findings can help us describe, quantify, measure, manage and predict future behavior of complex systems to achieve the highest rates of self organization. This method can be applied to Physical, Chemical, Biological, Ecological, Economical and other complex systems and in network theory. With a quantitative measure we can conduct exact scientific research on self-organization of complex systems and networks, progressive development, evolution and co-evolution, complexity, etc. We are working on applying the method to other systems from various fields from Physics and Chemistry to Economics and Cities and describe the processes, origin, mechanisms of self-organization.

ACKNOWLEDGMENTS

The author thanks Assumption College for support and encouragement of this research, and Prof. Slavkovsky, Prof. Schandel, Prof. Sholes, Prof. Theroux and Prof. Cromarty for discussion of the manuscript. The author thanks Francis M.